# Identifying interatomic potentials for the accurate modeling of interfacial segregation and structural transitions

Yang Hu[1], Jennifer D. Schuler[1], Timothy J. Rupert[1,2,*]

[1] Chemical Engineering and Materials Science, University of California, Irvine, California 92697, USA

[2] Mechanical and Aerospace Engineering, University of California, Irvine, California 92697, USA

* To whom correspondence should be addressed: trupert@uci.edu

**Abstract**

Chemical segregation and structural transitions at interfaces are important nanoscale phenomena, making them natural targets for atomistic modeling, yet interatomic potentials must be fit to secondary physical properties. To isolate the important factors that interatomic potentials must capture in order to accurately model such behavior, the performance of four interatomic potentials was evaluated for the Cu-Zr system, with experimental observations used to provide validation. While experimental results show strong Zr segregation to grain boundary regions and the formation of nanoscale amorphous complexions at high temperatures and/or dopant compositions, a variety of disparate behaviors can be observed in hybrid Monte Carlo/molecular dynamics simulations of doping, depending on the chosen potential. The potentials that are able to recreate the correct behavior accurately reproduce the enthalpy of mixing as well as the bond energies, providing a roadmap for the exploration of interfacial phenomena with atomistic modeling. Finally, we use our observations to find a reliable potential for the Ni-Zr system and show that this alloy should also be able to sustain amorphous complexions.

**KEYWORDS:** interfacial segregation; structural transition; complexions; molecular dynamics; interatomic potential

# 1. Introduction

Grain boundaries are important planar defects that influence a variety of material behaviors such as creep resistance, densification during sintering, fracture, fatigue, and thermal stability [1-6]. The structure and chemistry of grain boundaries can be altered by solute segregation, allowing material properties to be tuned and optimized by a method sometimes called "grain boundary segregation engineering" [7, 8]. Such a framework can be used to enhance the creep resistance of nanostructured alloys, as shown by Darling et al. [9] for nanostructured Cu-10 at.% Ta. These alloys contained Ta nanoclusters at the grain boundaries between Cu crystals and exhibited creep rates which were 6–8 orders of magnitude lower than most of reported data for other nanocrystalline metals. The concept of segregation engineering has also been applied to accelerating sintering by Park and Schuh [10], who observed the formation of Cr-rich phases between W-rich particles during the sintering of W-15 at.% Cr compacts. These Cr-rich regions acted as rapid transport pathways for W diffusion and thus reduced the sintering temperature and time needed for consolidation to full density. Raabe and coworkers [7, 11, 12] reported a nanoscale phase transformation from martensite to austenite due to Mn segregation in an Fe–9 at.% Mn maraging steel, showing that this interfacial phase transformation led to an increase of impact toughness. In addition to the examples of phase separation shown above, segregating dopants can also facilitate the formation of "complexions," or interfacial states whose existence depends on the adjacent crystals [13-15]. Dillon et al. [14] reported six types of discrete complexions, categorized by complexion thickness, in doped and undoped alumina. Khalajhedayati and coworkers [16, 17] applied this concept to nanostructured metals, where nanoscale amorphous complexions were observed in Cu-Zr alloys and these amorphous intergranular films dramatically improved the ductility of the material without sacrificing strength.

Most of the research studies mentioned above were in some way combined with atomistic modeling to explore the physical mechanisms responsible for the behavior observed in experiments. For instance, Darling and coworkers [9, 18] found that the bowing of grain boundaries at Ta clusters observed in experiments could be confirmed by molecular dynamics simulation and was consistent with a Zener pinning model. In addition, experiments can have limitations on what can be measured or contain a convolution of many factors that are difficult to separate. For example, Khalajhedayati et al. [16] showed that amorphous complexion were formed in nanocrystalline Cu-Zr but also found that complexion thickness varied significantly at different boundaries. Unfortunately, high resolution electron microscopy is limited to interfaces that can be viewed edge on, meaning a complete categorization of boundary structures in the material is extremely difficult, if not impossible, experimentally. However, Pan and Rupert [19] were able to provide a fundamental understanding of boundary-to-boundary variations in complexion thickness by building bicrystalline models for hybrid Monte Carlo/molecular dynamics simulations. These authors showed that different grain boundaries vary in their ability to collect Zr dopants and transition to an amorphous complexion at different global Zr compositions. These variations can explain the variety of boundary thicknesses seen in an experiment for a single global Zr composition. As another illustrative example, Frolov et al. [20] reported on interesting filled-kite and split-kite structures at $\Sigma 5$ (310) and $\Sigma 5$ (210) grain boundaries in pure Cu, predicting a transition between them under certain conditions which would be difficult to isolate experimentally.

While these examples demonstrate the value of atomistic modeling, these simulations are only as reliable as the potentials used to represent atomic interactions. Typically, the choice of interatomic potential is guided by experimental data and/or first-principles calculations. In some cases, the relevant material properties or defects energies of interest are obvious. For example, a direct fit to anisotropic elastic moduli can be used if elasticity is of interest [21], while fitting to the experimental phase diagram will allow precipitation and phase separation to be modeled [22]. In other cases, it is more difficult to isolate the essential properties to be fit for accurate simulation, particularly when a given phenomenon is not directly related to an equilibrium thermodynamic parameter in the bulk. For instance, Dziedzic et al. [23] focused on the structure and mass transport properties of liquid Al-Cu alloys and compared three distinct potentials by their ability of reproducing total and partial pair correlation functions, densities, angular distribution functions, coordination numbers, and self-diffusion coefficients. These authors found that two of the potentials failed to produce reasonable melting temperatures or densities of the targeted alloys, as well as missing the mark on other important properties, and thus were not appropriate for their study. However, even the potential which reproduced these properties still had difficulty reproducing local chemical ordering in an accurate manner. Similarly, Malerba et al. [24] tested the efficacy of four potentials for modeling radiation damage in Fe. These authors found that interatomic potentials fitted to first-principles forces (i.e., the slopes of potential-energy surfaces) or liquid structure factors (i.e., pair correlation functions) were able to more closely match available first-principles data for point defect formation and migration, key behaviors of interest when simulating radiation damage. However, for interfacial segregation and structural transitions, no guidelines currently exist for the selection of a suitable potential. The majority of the interesting mechanisms are occurring at the grain boundary region, yet most potentials are created using the equilibrium properties of bulk phases.

In this paper, we test different interatomic potentials for the Cu-Zr system with the goal of creating guidelines for potential selection to enable the accurate simulation of interfacial phenomena and structural transitions. Four interatomic potentials are chosen for this task, covering a range of functional forms and physical properties used for fitting. A set of new experimental observations on sputtered Cu-Zr thin films that are shown here, as well as past literature reports [16, 17], demonstrate that (1) Zr should exhibit a strong tendency to segregate to the grain boundaries and (2) amorphous intergranular films with stable nanoscale thicknesses should be created at high temperatures and high Zr concentrations. These experimental reports provide a baseline for comparison of the different potentials that are studied. We find that the functional form of the interatomic potential is less critical than the choice of properties chosen for fitting. Specifically, the enthalpy of mixing and the bond energies, must be faithfully represented to predict the chemical segregation and transition to an amorphous intergranular complexion. This can be done multiple ways, as one of the reliable potentials was created by fitting the ab initio atomic forces from a variety of atomic configurations while the other was created by calibrating to the liquid enthalpy of mixing and diffraction data from amorphous alloys. Armed with these guidelines, we extend our work to another promising system, Ni-Zr alloys, finding that amorphous films can be formed but that Zr only weakly segregates to the interfaces.

## 2. Methods

### 2.1. Experimental materials processing and characterization

Cu-Zr alloy films (composition of Cu-4.3 at.% Zr) were deposited onto Cu substrates with magnetron co-sputtering at a temperature of 400 °C and an Ar pressure of 1.5 mTorr in an Ulvac JSP 8000 metal deposition sputter tool. All samples were sealed under vacuum in high purity quartz tubes and annealed under vacuum at 500 °C for 24 h to promote grain growth and the segregation of dopants to the grain boundary, followed by a one-minute annealing treatment at 900 °C (0.92 $T_{solidus}$) to encourage grain boundary premelting. The samples were then rapidly quenched by being dropped into water to preserve any phases or grain boundary complexions occurring at high temperature. Transmission electron microscopy (TEM) samples were made from the films using the focused ion beam (FIB) lift-out technique [25] on an FEI dual beam Quanta 3D microscope using Ga+ ions. High resolution TEM (HRTEM) and energy-dispersive X-ray spectroscopy (EDS) were performed on an FEI Titan at 300 kV. Fresnel fringe imaging (FFI) was used to identify grain boundary films and to ensure edge-on orientation with respect to the grain boundary. A more detailed experimental study of complexion formation in a variety of Cu-rich films can be found in Ref. [26].

### 2.2. Computational methodology

Bicrystal models containing two Σ5 (310) grain boundaries were used as starting configurations. A Σ5 (310) grain boundary has a small, repeating kite-shaped structure (shown in Fig. 1) that allows for a reasonable simulation cell size, yet can act as a representative high-angle and high-energy grain boundary. Periodic boundary conditions were applied in all directions during the simulations. The simulation box has a length of approximately 23 nm (X direction), height of 11 nm (Y direction), and thickness of 4 nm (Z direction) containing 95,520 atoms. A hybrid Monte Carlo/molecular dynamics simulation method was used to simulate grain boundary segregation and any subsequent structural transitions [27]. This technique has been successfully used to model phase transformation and grain boundary segregation in earlier work. For example, Koju et al. [18] used Monte Carlo/molecular dynamics methods to study the Cu-Ta system and observed the formation of Ta-rich clusters at the grain boundary, a finding which was consistent with experimental observation in Cu-Ta [9]. Classical molecular dynamics simulations controlled the structural relaxation and were performed using the Large-scale Atomic/Molecular Massively Parallel Simulator (LAMMPS) package with an integration time step of 1 fs [28]. An isothermal-isobaric (NPT) ensemble with the Nose-Hoover thermostat/barostat was applied to relax samples at two temperatures (600 K and 1000 K) under zero pressure. To find chemical equilibrium, Monte Carlo simulations in a variance-constrained semi-grand canonical ensemble were performed after every 100 molecular dynamics steps.

Dopant concentrations of 0.4 and 4 at.% Zr were used to compare and contrast the four potentials for Cu-Zr, supplemented by a complete investigation of 0-10 at.% Zr for one of the potentials that as found to be reliable. For the later simulations of Ni-Zr, compositions between 0 and 10 at.% Zr were explored as well. While high temperature and high Zr boundary concentration promotes the formation of thicker disordered intergranular films, ordered grain boundaries have been reported at low temperatures and dopant concentrations [17]. After structural and chemical equilibrium were achieved, a conjugate gradient energy minimization was used to remove any remaining thermal noise. This quenching allowed the interfacial structure obtained during the doping process to be preserved and clearly analyzed. All structural analysis and visualization of atomic configurations used the open-source visualization tool OVITO [29], with the local crystal structure of each atom identified according to common neighbor analysis (CNA) [30]. Four interatomic potentials, two using the embedded-atom-method (EAM) formulation [31] and two using the Finnis-Sinclair formulation [32], were used. The potentials were developed by Cheng et al. [33], Ward et al. [34], Zhou et al. [35] and Mendelev et al. [36]. For brevity, we refer to these potentials using a combination of the first letter of each author's last name in the following sections. Thus, the potentials are referred to as the CSM potential, WAFW potential, ZJW potential, and MKOSYP potential, respectively.

Fig. 2 shows the potential energy, global and grain boundary concentration of Zr presented as a function of Monte Carlo step, of a bicrystal sample doped with 4 at.% Zr at 1000 K. In Fig. 2(a), the potential energy curve first has a very steep slope because the global dopant concentration is increasing to ~4 at.% Zr. This means that new dopants are rapidly being added to the system. After ~30 Monte Carlo steps, the global Zr composition has reached its target value and only fluctuates around this value by a small amount until the sample is equilibrated. In this regime, the existing Zr atoms are only switching sites, so the energy changes more slowly than before. A zoomed view (presented in the inset) of the potential energy curve near the end of Fig. 2(a) shows that that the potential energy keeps changing with additional Monte Carlo steps, meaning the system has not yet reached equilibrium. Figure 2(b) shows that the potential energy actually rises by a small amount by the end of the simulation. This increase is associated with Zr dopants oversaturating the boundary during the early stages of the simulations. As the simulation progresses, the boundary loses this extra Zr until an equilibrium configuration is found. Since a boundary film is forming in this sample, time is needed for an equilibrium film thickness and dopant concentration to be reached. The system is considered to be equilibrated when the absolute value of the fitted slope of the potential energy over the prior 1000 Monte Carlo steps is less than 0.001 eV step$^{-1}$.

## 3. Results

### 3.1. Experimental results for benchmarking

To provide an experimental baseline that can be used to validate the different potentials, Cu-Zr thin films were sputter deposited and subsequently processed to promote transformation of the grain boundary structure. While the average film composition was 4.3 at.% Zr, local variation in the Zr concentration at the grain boundaries was also observed, accompanied by disordered nanoscale intergranular films after quenching from 900 °C. Fig. 3(a) shows a HRTEM image of a grain boundary with an amorphous complexion that is approximately 2 nm thick. Fig. 3(b) is the corresponding EDS line profile scan showing that the concentration of Zr is ~1 at.% in grain interior and ~6 at.% at the grain boundary. The exact EDS composition values are subject to uncertainty due to spatial averaging from the electron beam interaction volume, but Zr segregation to the grain boundary is clearly shown. Khalajhedayati and coworkers [16, 17] studied the Cu-Zr system as well and found similar results. In that study, a mechanically alloyed Cu-3 at.% Zr powder was created by ball milling followed by annealing and quenching, and those authors found similar disordered nanoscale intergranular films when the annealing temperature was at or above 850 °C. In contrast, samples that were annealed at 550 °C or 750 °C only contained ordered grain boundaries. Similar reports of segregation and amorphous film formation in other materials have also be reported in the literature (see, e.g., [14, 26, 37]).

Detailed experimental studies of grain boundary complexions face unique challenges such as the difficulty of finding edge-on grain boundaries for inspection, grain overlap in nanocrystalline TEM samples that complicates data analysis, and spatial averaging of compositional measurements such as EDS across the grain boundary. While more detailed experimental observations on many boundaries would certainly be useful, such a study is beyond the scope of this paper and initial modeling attempts evaluated against available data can provide important benchmarks for future work. All of the experimental evidence presented here and available in the literature shows (1) interfacial segregation of Zr and (2) a structural transition to nanoscale amorphous complexions at high temperatures and sufficient dopant compositions. Therefore, interatomic potentials which are able to accurately describe interfacial segregation and grain boundary structural transition must be able to recreate these two features that were observed in the experiments.

### 3.2. Zr segregation and complexion formation simulated by different potentials

Although experiments give a consistent view of what is to be expected from Cu-Zr as boundaries are doped at elevated temperatures, the four potentials studied here predicted a variety of behaviors. Fig. 4-7 show the $\Sigma 5$ (013) grain boundary doped with either 0.4 at.% Zr or 4 at.% Zr at either 600 K or 1000 K, as obtained by the ZJW, WAFW, CSM, and MKOSYP potentials, respectively. The two compositions and temperatures were chosen to show the different type of complexions that are possible, ranging from monolayer segregation to amorphous intergranular films. In the image on the left side of each frame, chemical information is provided, where red atoms are Cu and blue atoms are Zr. In the images on the right, structural information is shown with face centered cubic atoms colored green, hexagonal close packed atoms colored red, body centered cubic atoms colored purple, icosahedral atoms colored yellow, and other atoms colored white.

We begin our analysis and discussion with the ZJW potential. Fig. 4(a) shows that Zr atoms segregate to the grain boundaries, occupying the kite tip at 600 K and 0.4 at.% Zr global concentration, while few can be found elsewhere. As the global concentration is increased to 4 at.% Zr shown in Fig. 4(b), crystalline order still exists and the kite-like structures can still be observed, but the boundary concentration has saturated and Zr begins to fill substitutional sites in the grain interior. When the temperature is increased to 1000 K at a low Zr composition, as shown in Fig. 4(c), the boundary appears thicker but the positions of Cu atoms at grain boundaries are not completely random. The kite-like structural units still exist to some extent, as shown by the black overlay, but the positions of Zr atoms are much more random within the boundary. Even so, the retention of the repeating structural units means that the boundary is still considered ordered. Finally, at a high Zr composition and high temperature, shown in Fig. 4(d), the positions of both Cu and Zr atoms at the grain boundary are randomly distributed and no repeating structural units are visible. The grain interior concentration at 4 at.% global Zr concentration is 3.6 at.% and 3.8 at.% at 600 K and 1000 K, respectively. These values are significantly higher than the ~1 at.% shown in Fig. 3(b), meaning the potential demonstrates a level of solid solubility that does not exist in real Cu-Zr alloys. Therefore, the ZJW potential overestimates the solubility of Zr in the Cu lattice and underestimates the segregation tendency of Zr atoms to grain boundary sites.

The WAFW potential was investigated next, again producing results (Fig. 5) that are quite different from experimental observations. An interesting feature that can be found is the formation of small Zr-rich clusters in samples doped with 0.4 at.% Zr at both 600 K and 1000 K, as shown in Fig. 5(a) and (b). Fig. 5(c) and (d) show that the formation of disordered intergranular films is captured at both temperatures for samples doped with 4 at.% Zr and the grain boundary concentrations are 35.6 at.% Zr at 600 K and 28.4 at.% Zr at 1000 K, respectively. Although the films are structurally amorphous as measured by the CNA parameter, they are chemically ordered (i.e., there is a patterning of chemical composition) and the grain boundary composition values are significantly higher than what was observed in the experimental reports. This high Zr segregation tendency and also the formation of Zr-rich clusters are signs that the potential has limitations for modeling complexion transitions. The disordering behavior observed here happens when the Zr-rich clusters at the grain boundaries grow larger and eventually connect to each other. However, no such clustering is observed experimentally.

While the two potentials discussed above do not accurately reproduce Zr segregation, the CSM and MKOSYP potentials faithfully reproduce the expected segregation and formation of nanoscale amorphous complexions in a manner that is similar to experimental observations. For the CSM potential, Fig. 6(a) shows that a monolayer complexion forms at the grain boundary with Zr atoms occupying the tip of the kite structure. Even at high temperature, as shown in Fig. 6(c), the grain boundary stays as ordered as the clean grain boundary in pure Cu as long as the Zr concentration is very low. As the global Zr concentration goes to 4 at.%, the grain boundary becomes fully disordered and a complexion of nanoscale thickness forms at both temperatures, with the higher temperature giving a much thicker amorphous film. This temperature dependence, which contrasts with the nearly constant thickness observed for the WAFW potential, is also observed in thermodynamic treatments of amorphous complexion formation [38]. The MKOSYP potential shows a similar transition in Fig. 7 from a monolayer complexion at low Zr concentration and temperature to an amorphous intergranular film at high Zr concentration and temperature.

Zr concentration profiles across the grain boundary are plotted in Fig. 8 to assess the grain boundary segregation tendency of each potential in a more quantitative manner and to make a closer comparison with experimental data. Strong grain boundary segregation occurs for the CSM, WAFW, and MKOSYP potentials, while the grain boundary modeled with the ZJW potential shows very little enrichment. The ZJW potential also demonstrates a relatively large solubility of Zr into the Cu lattice that is not consistent with the experimental observations. Fig. 8 does show that the MKOSYP potential gives a higher grain boundary concentration and a thinner grain boundary film than the CSM potential. One possible explanation for this observation is that the MKOSYP potential gives a higher melting temperature for Cu than the CSM potential (melting temperature of 1355 K for the MKOSYP potential and 1175 K for the CSM potential). The melting temperature of Cu was obtained for each potential using a coexistence simulation following Allen and Tildesley [39], where the NPT ensemble was applied at different temperatures and under zero pressure while the interface between a solid and liquid phase was tracked. Since all of our simulations were run at the same two temperatures, variations in melting point mean that different potentials can have small variations in the homologous temperature ($T/T_{melting}$) for a given simulation. Interfacial segregation behavior is strongly influenced by temperature and higher homologous temperatures tend to lead to thicker amorphous intergranular films, which is consistent with our observations (i.e., 1000 K is a higher homologous temperature for the CSM potential). In the end, even though there are noticeable differences in the exact grain boundary and grain interior compositions for the MKOSYP and CSM potentials, both capture the important segregation and transformation phenomena demonstrated by the experimental study.

To give an overall view of the complexion transition process for the CSM potential (Pan and Rupert provided a complete study using the MKOSYP potential in [19]), the variation of film thickness as well as grain boundary concentration with increasing global concentration are plotted in Fig. 9. At low global Zr compositions, an ordered, monolayer complexion is formed at the grain boundary. The grain boundary thickness remains constant and very small as global composition increases, while the grain boundary concentration increases rapidly. After the global composition reaches ~0.5 at.% Zr, a gradual transition stage begins, where some parts of the grain boundary structure become partially disordered while the rest of the boundary remains ordered. This manifests as an increase in the average film thickness. For the 1000 K sample, the film thickness grows more quickly than the amount of Zr atoms segregating to the grain boundaries, resulting in a drop in grain boundary concentration at a global composition of 0.8 at.% Zr. A similar, temporary drop in grain boundary composition can be found in the 600 K sample at a global concentration of ~2.5 at.% Zr. In the nanoscale amorphous film and wetting film stages, the grain interior concentration is very low for both temperatures, reaching only about 0.01 at.% at 600 K and 0.1 at.% at 1000 K.  The observed value for 1000 K is comparable to the reported solubility of 0.12 at.% Zr in Cu at 972 °C (1245.15 K) [40].  In these stages, due to the complete disordering of the grain boundary structure and the rapid increase in grain boundary thickness, many more possible segregation sites are available for Zr atoms.  A nanoscale film is a true complexion whose chemistry and structure depends on the abutting crystals, so the grain boundary Zr concentration continues to increase as global Zr composition increases.  Alternatively, a wetting film can be distinguished based on the fact that it is a true bulk phase and therefore has a constant grain boundary concentration, as indicated by the blue dashed lines in Fig. 9(b), both curves reach a plateau eventually.  Pan and Rupert [19] also reported the saturation of grain boundary concentration of Zr at both temperatures.

## 4.  Discussion

To understand why only certain potentials can faithfully recreate grain boundary segregation and complexion transitions, a discussion of the important physical parameters responsible for this behavior is necessary.  Grain boundary segregation is generally governed by the competition between the Gibbs free energy of a system with dopants in the bulk and the same system with dopants at the interface, under a given set of thermodynamic conditions such as temperature, pressure and chemical potential.  The Gibbs free energy for segregation can be expressed as $\Delta G_{seg} = \Delta H_{seg} - T\Delta S_{seg}$, where $\Delta H_{seg}$ is the enthalpy of segregation, $T$ is temperature, and $\Delta S_{seg}$ is the entropy of segregation.  Ordered grain boundaries were observed at low temperatures, where the entropic contribution to the total energy will be small and the focus can remain on the enthalpic term.   In the discussion that follows, a negative $\Delta H_{seg}$ means segregation to grain boundary while a positive value indicates grain boundary depletion.   It is important to note that different models for calculating $\Delta H_{seg}$ might use different sign convention.   For example, Murdoch and Schuh [41] calculated the enthalpy of mixing for a large number of transition metal alloys, but used a definition of the enthalpy of segregation where positive values denoted segregation.

Here, we use the Wynblatt-Ku [42, 43] model as a basis for the discussion of ordered grain boundary segregation at low temperatures.   This model considers three main contributions to interfacial segregation: (1) the elastic contribution, (2) the interfacial energy contribution, and (3) the interatomic contribution.   The total formulation for segregation enthalpy is shown in Eq. (1):

$$\Delta H_{seg} = -\frac{24\pi B \mu r_I r_M (r_I - r_M)^2}{3Br_I + 4\mu r_M} + (\sigma_I - \sigma_M)A^\Phi - \frac{2\Delta H_m}{ZX_I X_M}[Z_L(X_I^\Phi - X_M) + Z_P(X_I - \frac{1}{2})] \quad (1)$$

where $r_I$ and $r_M$ are the atomic radii for the solute and solvent, $B$ is the bulk modulus of solute I, $\mu$ is the shear modulus of solvent M, $\sigma_I$ and $\sigma_M$ are the interfacial energies for elements I and M, $A^\Phi$ is the interface area per atom, $X_I$ and $X_M$ are the bulk concentration of I and M respectively, $X_I^\Phi$ is the interfacial concentration of the solute, $\Delta H_m$ is the enthalpy of mixing of the M-I alloy, $Z$ is the coordination number, $Z_L$ is the number of lateral bonds made by an atom within its plane, and $Z_P$ is the number of bonds made with adjacent planes of atoms.

While all of the parameters in Eq. 1 can affect an alloy's segregation enthalpy, many can be ruled out as the cause behind the disparate observations from our multiple simulations.   For example, for a given grain boundary type, such as the Σ5 (310) boundary studied here, $Z$, $Z_L$, and $Z_P$ will be the same. In addition, parameters such as $r_I$, $r_M$, and $A^\Phi$ come directly from the lattice constants of Cu and Zr, which are relatively easy targets that any reasonable potential should reproduce. The calculated lattice constants for Cu and fcc-Zr are shown in Table 1 for the four potentials studied here, along with experimental data or density functional theory (DFT) calculations to provide a reference. The lattice parameters of pure Cu and Zr were obtained from Ref. [44] and come from DFT calculations using the projector augmented wave approach and the Perdew-Burke-Ernzerhof exchange-correlation generalized gradient approximation (GGA) functional. These values were obtained from the database of the Materials Project (www.materialsproject.org). A second measurement for the lattice parameter of Cu comes from Ref. [45], which was obtained experimentally at room temperature using the asymmetric film method. All of the potentials accurately reproduce the equilibrium lattice constants to an accuracy of approximately 0.5% as compared to the reference values. Other parameters such as $\Delta H_m$ of the alloy system are more complicated and harder to recreate with the cross-potential terms, therefore being likely sources of poor fitting.

To probe the ability of the four potentials to reproduce the enthalpy of mixing of the Cu-Zr system, a 7.23 nm × 7.23 nm × 7.23 nm pure Cu model was first generated, and the Monte Carlo/molecular dynamics hybrid method was then used to introduce Zr into this bulk sample. An NPT ensemble was applied to the system at 1873 K and zero pressure, with the high temperature chosen so that our calculations could be compared with available experimental data taken from the liquid phase. The $\Delta H_m$ was calculated according to Eq. (2) [46]:

$$\Delta H_m = H_{A-B} - (X_A H_A^0 + X_B H_B^0) \qquad (2)$$

where $H_{A-B}$ is the molar enthalpy of A and B binary solution, $H_A^0$ and $H_B^0$ are the standard molar enthalpies of components A and B, and $X_A$ and $X_B$ are the mole fractions of component A and B. The units of these calculated values were changed to eV/atom. Experimental data from Turchanin et al. [47] is used as a reference. These authors first obtained the enthalpy of mixing in the concentration interval of 0-54 at.% Zr using a high-temperature isoperibolic calorimeter and then converted these measurements to partial molar enthalpies of mixing. Extrapolation was then used to establish the equation of integrated enthalpy of mixing which can be applied for the entire concentration range from 0-100 at.% Zr. As shown in Fig. 10, the CSM and MKOSYP potentials are able to reproduce the variation of enthalpy of mixing with respect to Zr concentration that is seen in the experimental data [47]. Alternatively, the enthalpy of mixing values obtained by the WAFW potential are positive at low Zr concentrations, marking a clear deviation from real material behavior. In addition, the enthalpy of mixing obtained by the ZJW potential is much more negative (approximately 400% lower) than the reported data.

Using the Wynblatt-Ku model introduced above and keeping all other variables constant, a positive $\Delta H_m$ will enhance segregation and clustering of like atoms. The positive values of enthalpy of mixing given by the WAFW potential contributes to the formation of Zr-rich clusters at the grain boundaries. On the other hand, if $\Delta H_m$ is much too negative, like the behavior produced by the ZJW potential, there will be very little segregation of Zr to the grain boundaries because Cu-Zr bonds are favored in the lattice. To understand these observations, one can return to inspect the fitting process used for each of the four potentials. For the ZJW potential, the pairwise cross-function is constructed solely based on the elemental two-body pair potentials, which use lattice constants, cohesive energies, bulk moduli, Voigt-average shear moduli, unrelaxed vacancy formation energies of pure Cu and Zr, and the dilute-limit heats of solution of Zr in Cu to fit the potential [35, 48]. Similarly, the WAFW potential is also constructed using the existed monatomic potentials from literature. Mixing enthalpies, bulk moduli, and lattice parameters of possible $B_2$ and $L1_2$ Cu-Zr intermetallics are used for fitting the cross-potential term in this case [34]. Due to these limited fitting procedures, the ZJW and WAFW potentials are, at best, only able to reproduce the enthalpy of mixing at certain Zr concentration (e.g., Cu doped with an extremely small amount of Zr), which limits their application. In addition, only a few crystalline structures are considered for fitting the cross-potential terms for these two interatomic potentials. Such a limited procedure is problematic when studying systems containing defects like grain boundaries or systems which undergo interfacial structural transitions to amorphous complexions that are structurally and compositionally complex.

In contrast, the MKOSYP potential is constructed based on a previous developed interatomic potential for Cu-Zr [49] but adds additional complexity associated with a new pairwise cross-function created using experimental diffraction data, the liquid density at 1500 K of amorphous $Cu_{64.5}Zr_{35.5}$, and the enthalpy of mixing of the liquid state [36]. With features of multiple crystalline and amorphous states incorporated into the fitting procedure, a better reproduction of interfacial segregation is achieved. The CSM potential incorporates additional complexity by using a force-matching method [50], where the potential energies, atomic forces (slope of potential-energy surfaces), and stress tensors of 954 configurations calculated by first-principles method are included in the fitting database. The addition of this detail allows the potential to more accurately describe the chemical bonding between atoms, since forces acting on atoms come from the interaction between electrons and the interaction between electrons and nuclei [50], all of which are captured by first-principles calculations. Moreover, a very broad range of structural configurations of pure Cu, pure Zr, and binary Cu-Zr was alloys were used for fitting, including liquid phases, metallic glasses quenched at different cooling rates, and crystalline structures with interfacial and point defects [33, 51]. These configurations contain a variety of bonding types associated with the entire gambit of material defects that are important for reproducing interfacial segregation.

In addition to segregation, the formation of amorphous complexions should also be related to a potential's ability to accurately reproduce the enthalpy of mixing. In their thermodynamic model for disordered complexion formation, Luo et al. [38] proposed that the stabilization of an amorphous intergranular film occurs when the free energy penalty for the formation of the film is smaller than the reduction in interfacial energy as a result of replacing a clean grain boundary with two crystal-liquid interfaces. A key consideration for the discovery of good glass formers is a negative enthalpy of mixing (see, e.g., [52]), meaning this parameter should be related to the energetic of the amorphous phase. In fact, Schuler and Rupert [26] recently created a set of materials selection rules for alloys that could sustain amorphous intergranular films where a negative enthalpy of mixing was also of primary importance. Thus, the accurate reproduction of enthalpy of mixing by the CSM and MKOSYP potentials also leads to a realistic picture of disordered complexion formation at high temperatures and dopant compositions.

Grain boundary energy can provide an additional metric to facilitate a comparison between the various potentials. The calculated energies of a clean $\Sigma 5$ (310) grain boundary simulated by each potential are shown in Table 1, along with a reliable energy value taken from DFT calculations using the projector augmented wave approach and the Perdew-Burke-Ernzerhof exchange-correlation GGA functional [53]. Both the CSM potential and the MKOSYP potential recreate grain boundary energy with errors of less than 3% compared to the baseline value, while the ZJW and WAFW potentials give energies that are much too low. For the CSM potential, fitting to ab initio atomic forces in a variety of material configurations such as liquid phases and crystal structures with point defects and interfacial defects appears to better capture the nuances of bonding in the grain boundary region. For the MKOSYP potential, the use of fitting to diffraction data for an amorphous Cu-Zr alloy is advantageous. An amorphous structure contains both free volume and a statistical distribution of bond lengths that are larger than the equilibrium lattice constant, both of which are features of a grain boundary's equilibrium structure. Thus, fitting a potential to either the atomic forces from first-principles or diffraction patterns of liquid or amorphous phases improves its ability to reproduce interfacial energy. In summary, the CSM and MKOSYP potentials provide overall better reproduction of the physical properties such as enthalpy of mixing and grain boundary energy that appear in Eq. (1) above, leading to more realistic modeling of interfacial segregation behavior.

## 5. Extension to the Ni-Zr system

To show that the observations above are useful for other materials besides Cu-Zr, we simulate interfacial segregation and complexion transitions in Ni-Zr. Schuler and Rupert [26] have recently shown that this alloy can sustain thick amorphous intergranular films, due to the segregation of Zr to the grain boundaries and the negative enthalpy of mixing, which reduces the penalty for an amorphous phase. The same bicrystal configuration was used here, containing two Σ5 (310) grain boundaries. An interatomic potential which fits to the ab initio atomic forces and thus satisfies the requirement of accurately reproducing the enthalpy of mixing (as shown in Fig. 11), bond energies and other physical properties was chosen [54]. A few examples of equilibrium grain boundary structures are provided in Fig. 12, with chemical information in the left panel of each figure part and structural information in the right panel. Figs. 12(a) and (b) show that of Zr occurs but that the lattice can also accept some Zr atoms. A sudden increase of film thickness occurs when the global concentration reaches 9 at.% Zr, signaling the formation of an amorphous nanoscale complexion. The spatial variation of Zr composition within the simulated sample is plotted in Fig. 13. The grain boundary region is enriched with Zr for all three samples, consistent with experimental reports [26]. The lower segregation tendency of Zr in Ni, as compared to Cu, is consistent with a higher solubility on the bulk phase diagram [56].

## 6. Conclusions

In this paper, we have identified the key characteristics of interatomic potentials which are able to accurately recreate interfacial segregation and complexion transitions. Four interatomic potentials were tested on their ability to recreate key experimental observations in the Cu-Zr system. Based on these results, the following conclusions can be made:

- Zr atoms segregate to Cu grain boundaries, eventually leading a transformation to nanoscale amorphous complexions as doping concentration and temperature increases.

- The ZJW potential underestimates the segregation tendency of Zr, while the WAFW potential captures the segregation of Zr but produces an erroneous clustering of Zr atoms in the boundary. In contrast, both the CSM potential and the MKOSYP potential are able to produce simulation result which mimic experimental observations.
- Accurate modeling of interfacial segregation and structural transitions requires the reproduction of physical quantities such as enthalpy of mixing and bond energies. To obtain precise values of these parameters, experimental or first-principles data from multiple phases and material states should be included in the fitting database used to create the cross-potential terms.
- Accurate simulations can also be performed on the Ni-Zr system, which demonstrates a more subtle tendency for grain boundary segregation of Zr but the ability to form amorphous intergranular films.

By understanding what features are needed for accurate modeling of interfacial phenomena, this work opens the door for further materials discovery. Alloys of interest can be tested in simple bicrystal models prior to experimentation, to identify promising material combinations that can sustain disordered complexions.

**Acknowledgements**

This research was supported by U.S. Department of Energy, Office of Basic Energy Sciences, Materials Science and Engineering Division under Award No. DE-SC0014232.

Table 1

| Element | Cu | | | | | | fcc-Zr | | | | |
|---|---|---|---|---|---|---|---|---|---|---|---|
| Interatomic Potential | ZJW[35] | WAFW[34] | CSM[33] | MKOSYP[36] | Reported data | ZJW[35] | WAFW[34] | CSM[33] | MKOSYP[36] | Reported data |
| Lattice Parameter (Å) | 3.615 | 3.614 | 3.597 | 3.639 | 3.616[44] 3.615[45] | 4.532 | 4.531 | 4.539 | 4.545 | 4.537[44] |
| Σ5 (310) GB energy (J/m²) | 0.783 | 0.784 | 0.901 | 0.903 | 0.88[53] | - | | | | |

Note: lattice parameters from Ref. [44] were obtained by DFT calculation using the projector augmented wave approach and the Perdew-Burke-Ernzerhof exchange-correlation GGA functional and was stored in the database of the Materials Project (www.materialsproject.org). Data from Ref. [45] was obtained experimentally at room temperature using the asymmetric film method. The grain boundary energy data from Ref. [53] was also obtained by DFT calculation using the projector augmented wave approach and the Perdew-Burke-Ernzerhof exchange-correlation GGA functional.

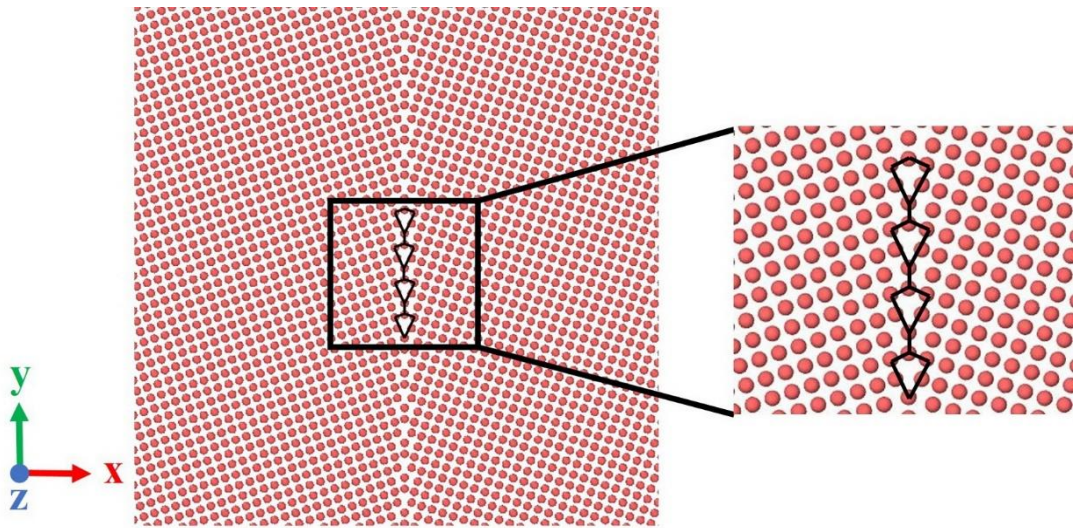

Fig. 1. A Σ5 (310) grain boundary in pure Cu (i.e., a clean grain boundary). The repeating kite-shaped structural unit is outlined by black lines.

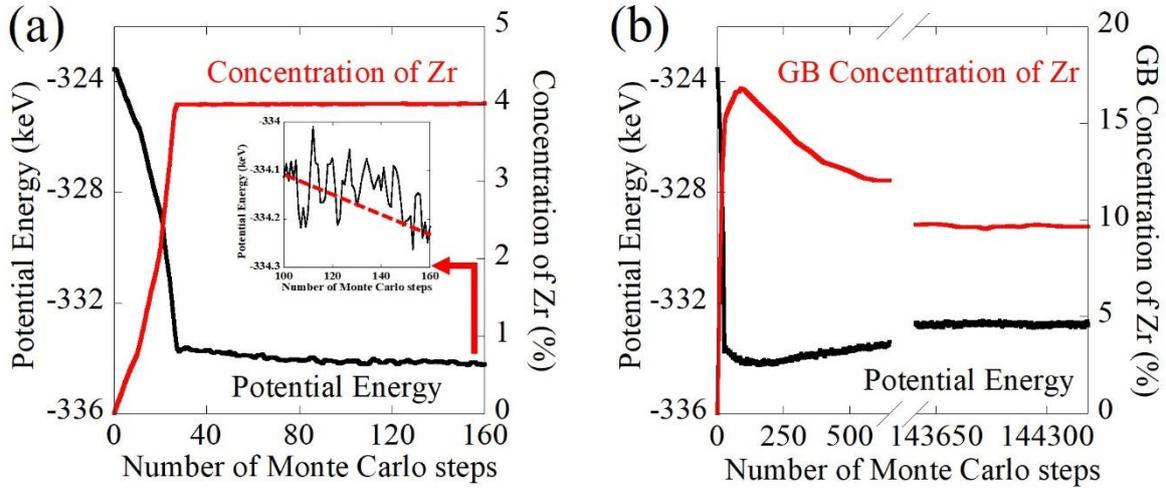

Fig. 2. (a) The variation of potential energy and global concentration of Zr with respect to Monte Carlo steps in the early stages of a doping simulation for a Cu sample with 4 at.% Zr at 1000 K. (b) The variation of potential energy and grain boundary concentration of Zr with respect to Monte Carlo steps for a Cu sample doped with 4 at.% Zr at 1000 K. The inset in (a) shows a zoomed view from 100-160 Monte Carlo steps and the red dashed line represents the average value of potential energy. In part (a), the slope of the curve changes at ~30 Monte Carlo steps because the global dopant concentration reaches its target of ~4 at.% Zr. In part (b), the potential energy demonstrates a small increase as a temporary oversaturation of the grain boundary is corrected.

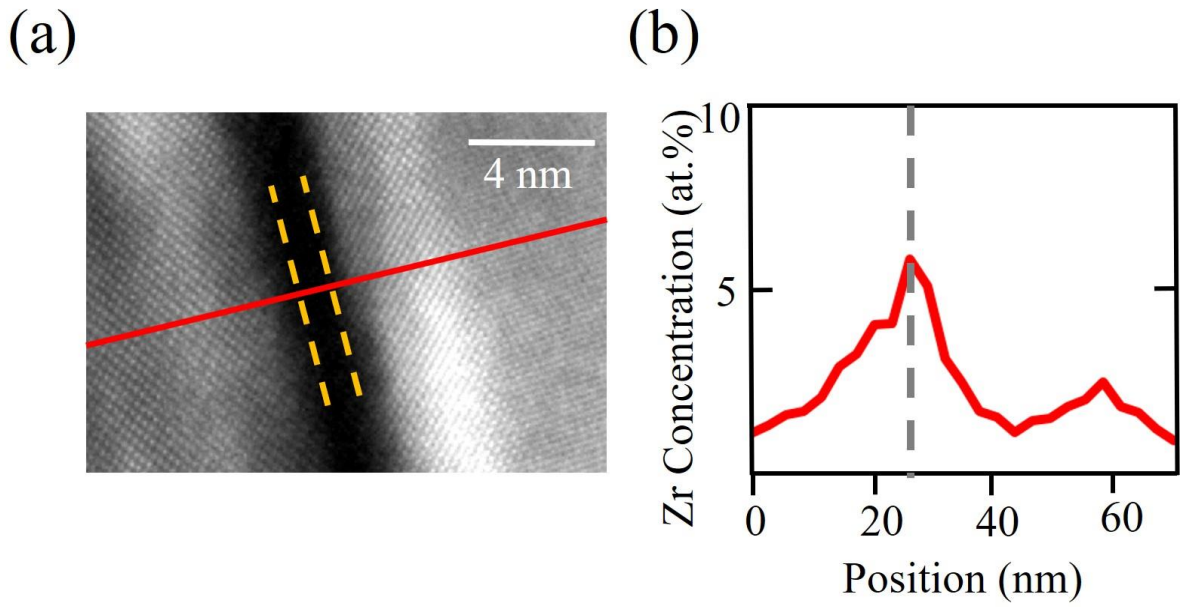

Fig. 3. (a) High resolution TEM image of a nanoscale amorphous complexion at the grain boundary in a sputtered Cu-Zr sample. (b) EDS line profile scan across the grain boundary, showing Zr enrichment at the interface. The red line in (a) gives the scan path while the dashed yellow lines roughly outline the amorphous film. The grey line in (b) denotes the grain boundary location.

ZJW Potential

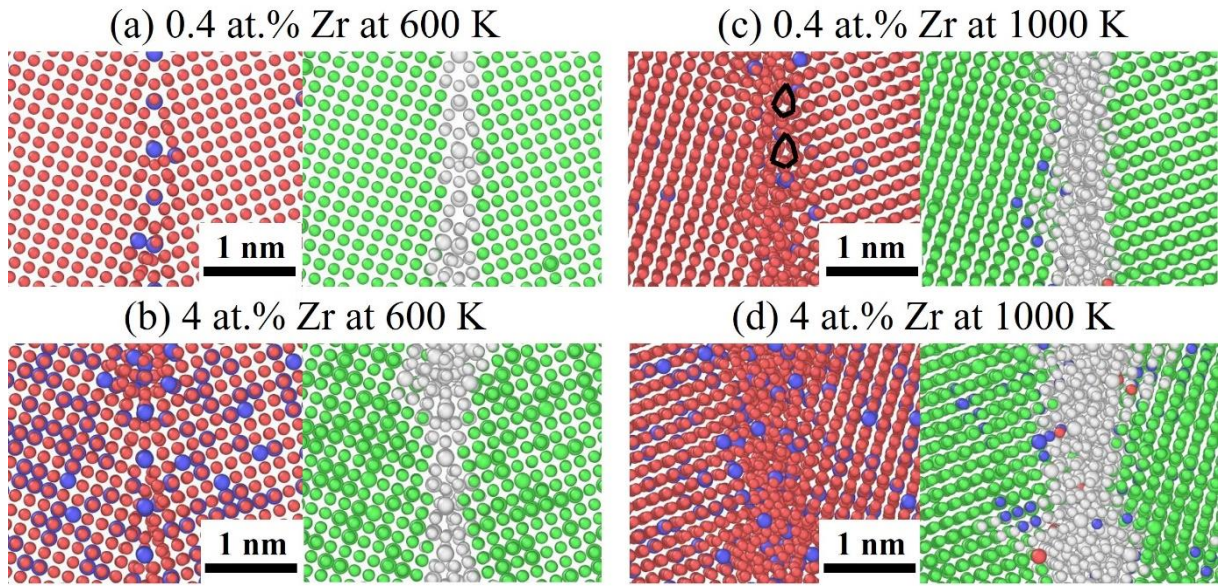

Fig. 4. The equilibrium chemical (left panel) and structural (right panel) information of the Σ5 (013) grain boundary in Cu doped with (a) 0.4 at.% Zr at 600 K, (b) 4 at.% Zr at 600 K, (c) 0.4 at.% Zr at 1000 K, and (d) 4 at.% Zr at 1000 K using the ZJW potential.  In the panels on the left, Cu atoms are colored red and Zr atoms are colored blue. In the panels on the right, face centered cubic atoms are colored green, hexagonal close packed atoms red, body centered cubic atoms purple, icosahedral atoms yellow, and other atoms white.

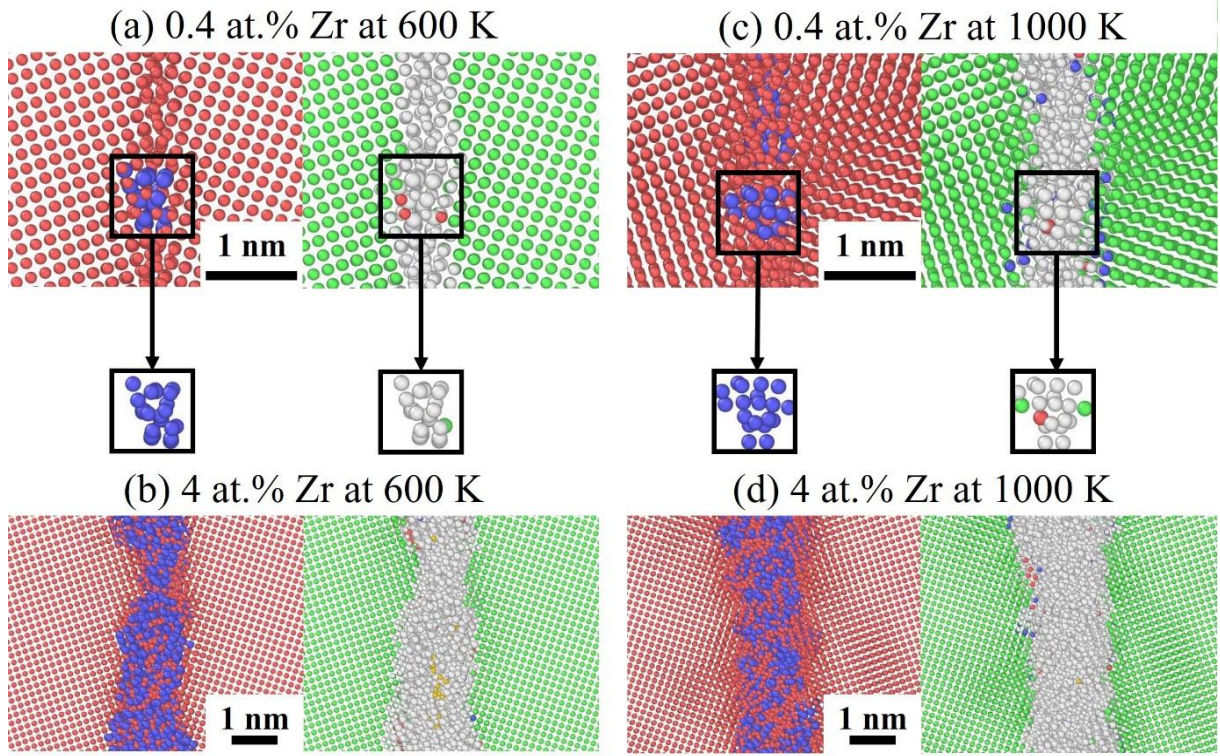

Fig. 5. The equilibrium chemical (left panel) and structural (right panel) information of the Σ5 (013) grain boundary in Cu doped with (a) 0.4 at.% Zr at 600 K, (b) 4 at.% Zr at 600 K, (c) 0.4 at.% Zr at 1000 K, and (d) 4 at.% Zr at 1000 K obtained with the WAFW potential. In the panels on the left, Cu atoms are colored red and Zr atoms are colored blue. In the panels on the right, face centered cubic atoms are colored green, hexagonal close packed atoms red, body centered cubic atoms purple, icosahedral atoms yellow, and other atoms white. The enlarged insets show Zr clustering within the boundary.

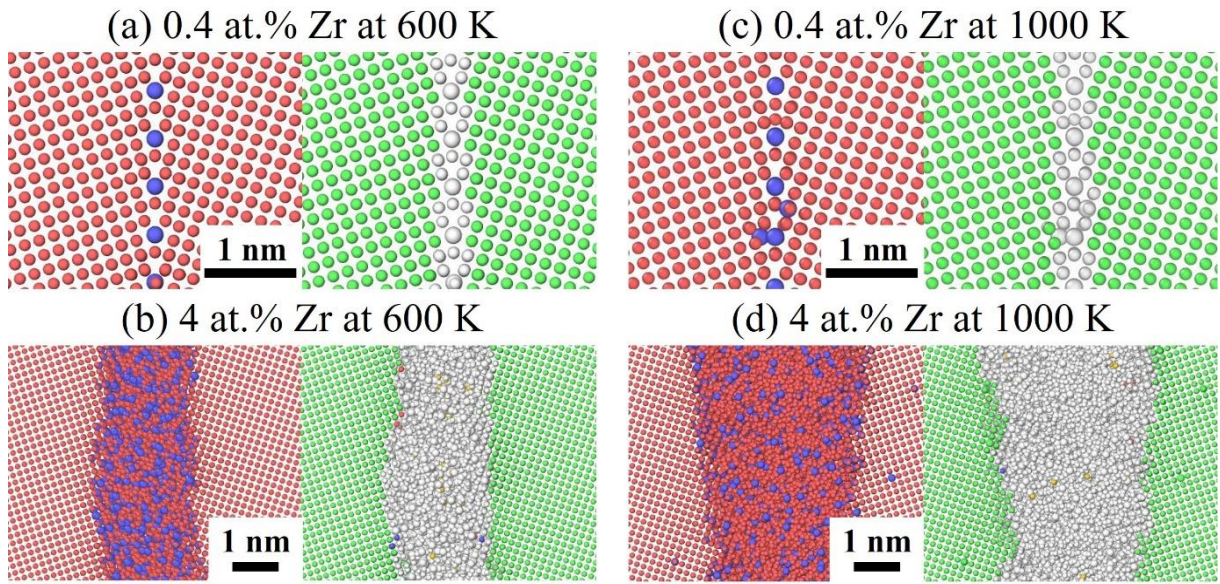

Fig. 6. The equilibrium chemical (left panel) and structural (right panel) information of the Σ5 (013) grain boundary in Cu doped with (a) 0.4 at.% Zr at 600 K, (b) 4 at.% Zr at 1000 K, (c) 0.4 at.% Zr at 600 K, and (d) 4 at.% Zr at 1000 K obtained using the CSM potential. In the panels on the left, Cu atoms are colored red and Zr atoms are colored blue. In the panels on the right, face centered cubic atoms are colored green, hexagonal close packed atoms red, body centered cubic atoms purple, icosahedral atoms yellow, and other atoms white.

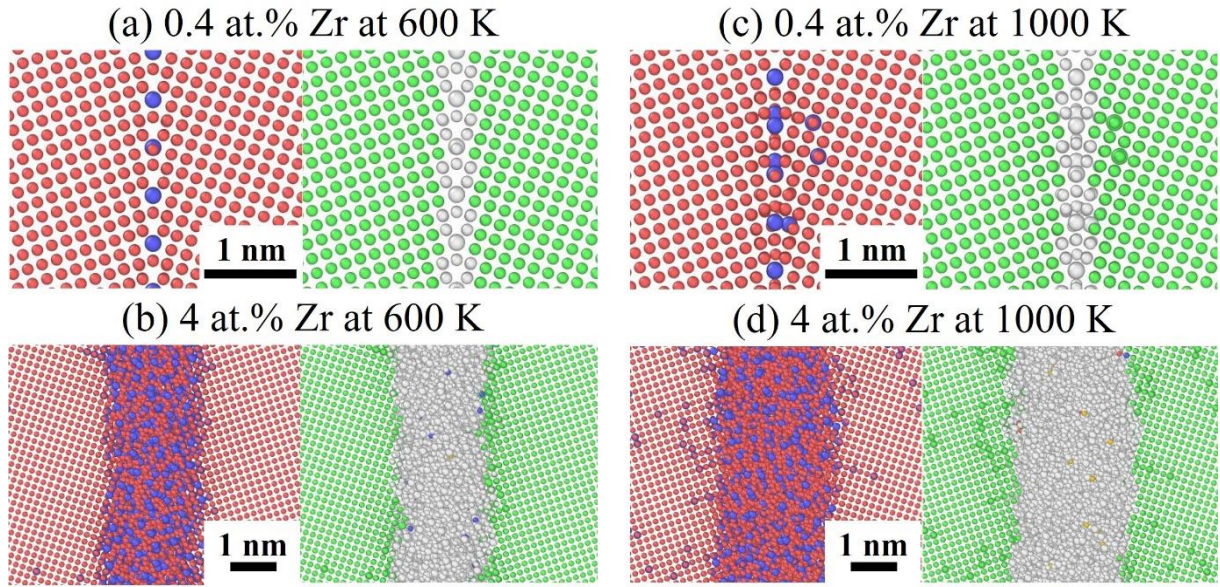

Fig. 7. The equilibrium chemical (left panel) and structural (right panel) information of the Σ5 (013) grain boundary in Cu doped with (a) 0.4 at.% Zr at 600 K, (b) 4 at.% Zr at 1000 K, (c) 0.4 at.% Zr at 600 K, and (d) 4 at.% Zr at 1000 K obtained using the MKOSYP potential. In the panels on the left, Cu atoms are colored red and Zr atoms are colored blue. In the panels on the right, face centered cubic atoms are colored green, hexagonal close packed atoms red, body centered cubic atoms purple, icosahedral atoms yellow, and other atoms white.

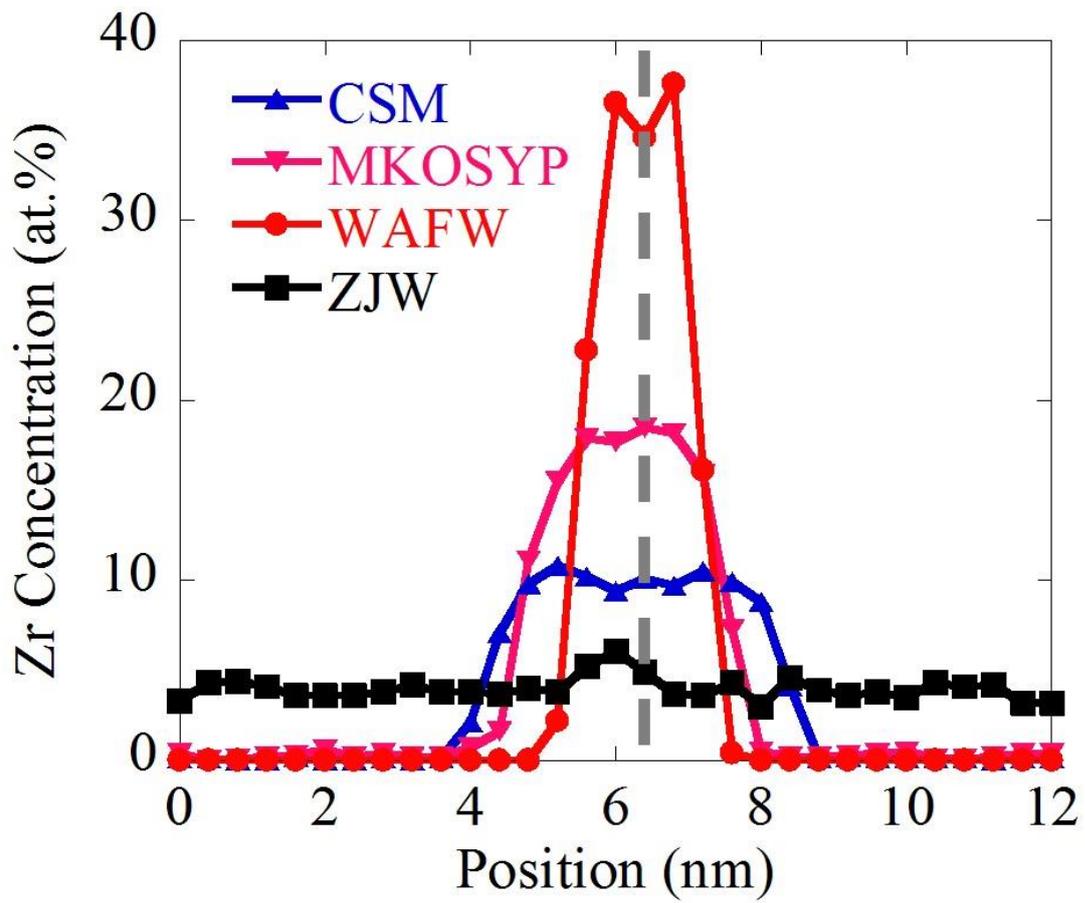

Fig. 8. Zr concentration profiles across the grain boundary in Cu samples doped with 4 at.% Zr at 1000 K simulated by four different interatomic potentials.

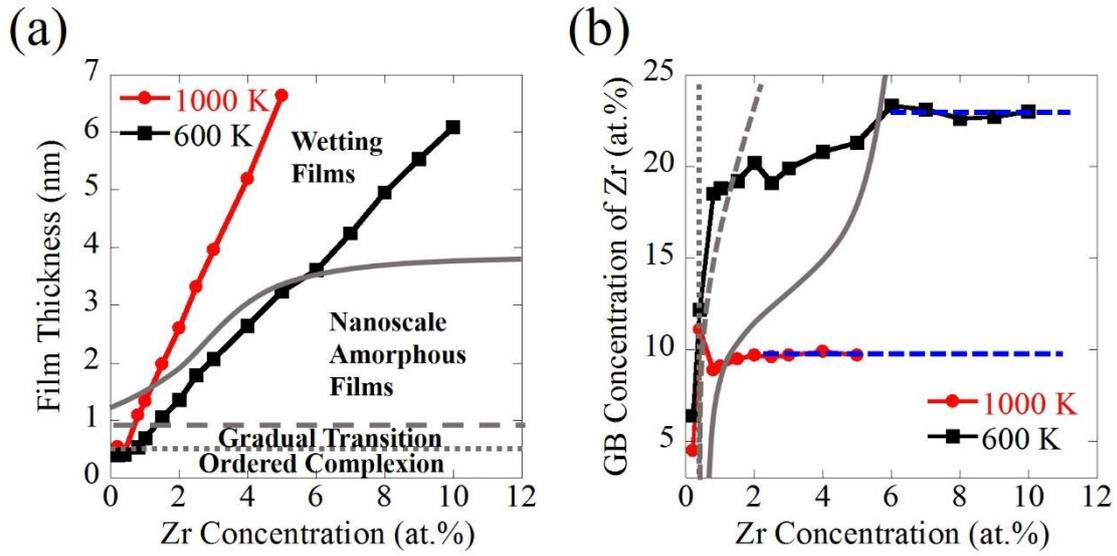

Fig. 9. (a) Film thickness and (b) grain boundary (GB) concentration of Zr with increasing global Zr concentration, as simulated by the CSM potential. Different complexion regions are marked in (a) and (b), while the dashed blue line in (b) shows the saturation of grain boundary concentration.

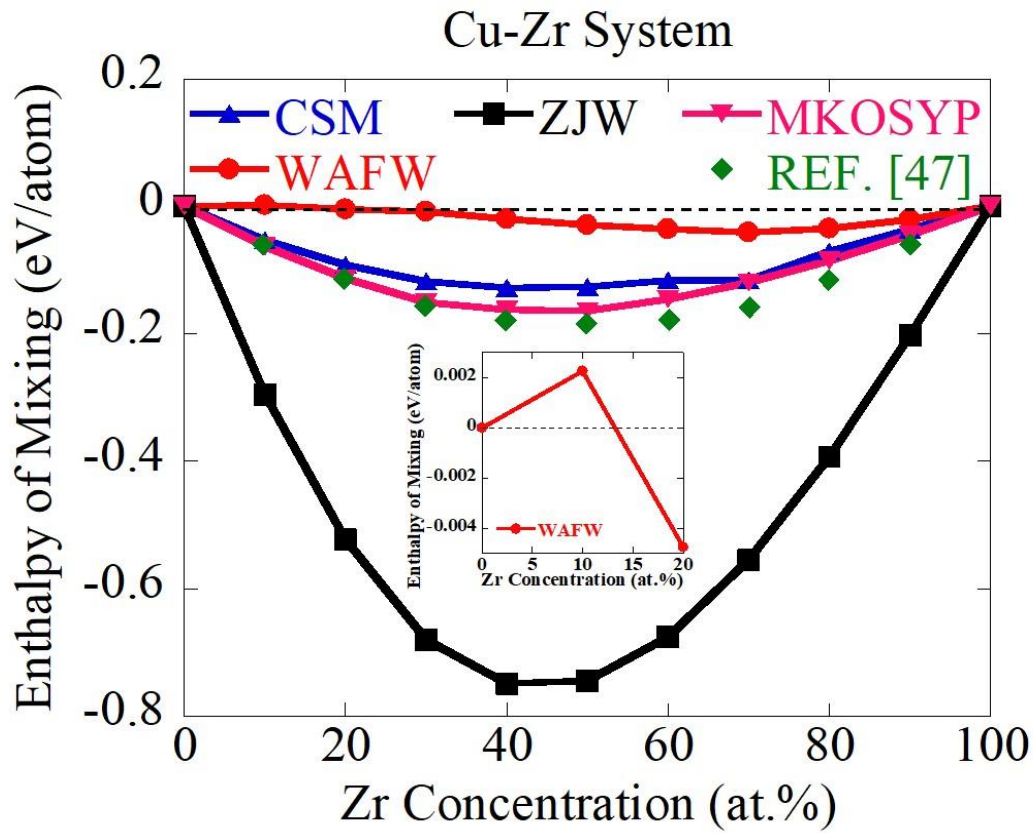

Fig. 10. Enthalpy of mixing of the Cu-Zr system calculated at 1873 K. The reference data comes from Turchanin [47]. The inset shows a zoomed view of the enthalpy of mixing from 0 at.% to 20 at.% Zr simulated by the WAFW potential.

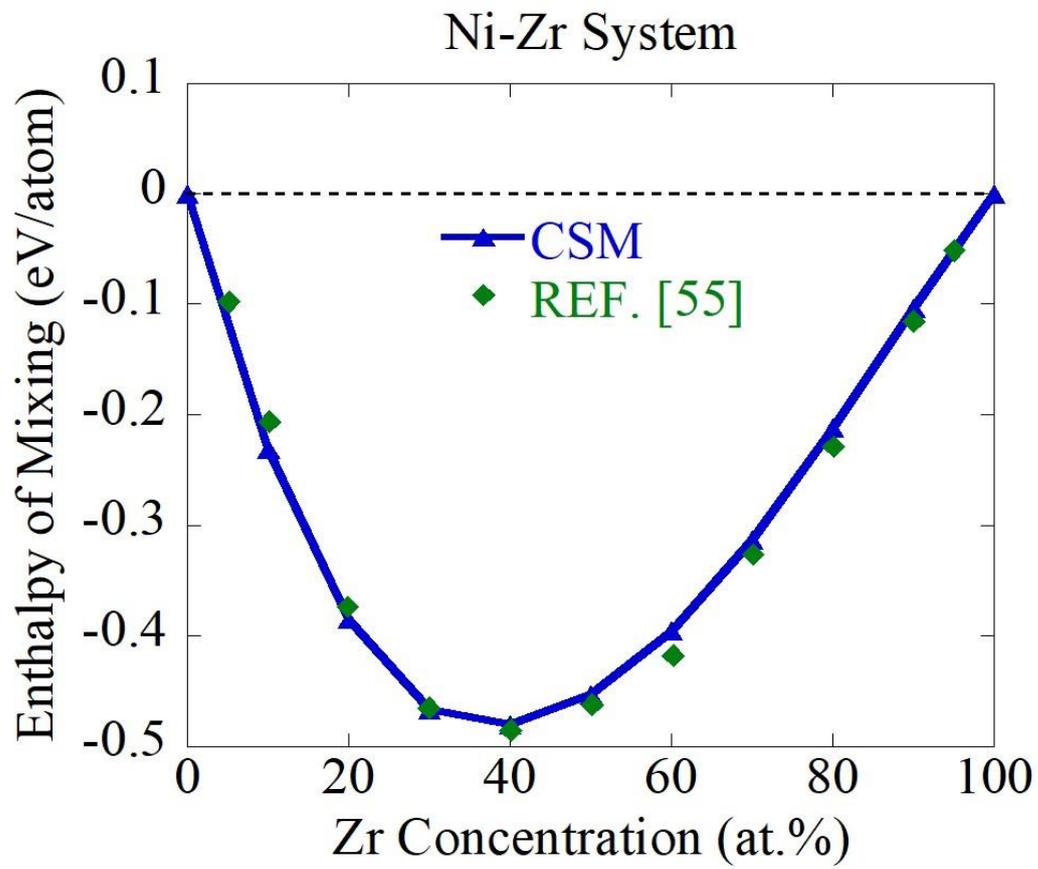

Fig. 11. Enthalpy of mixing of Ni-Zr system calculated at 1873 K. The reference data comes from Turchanin et al. [55].

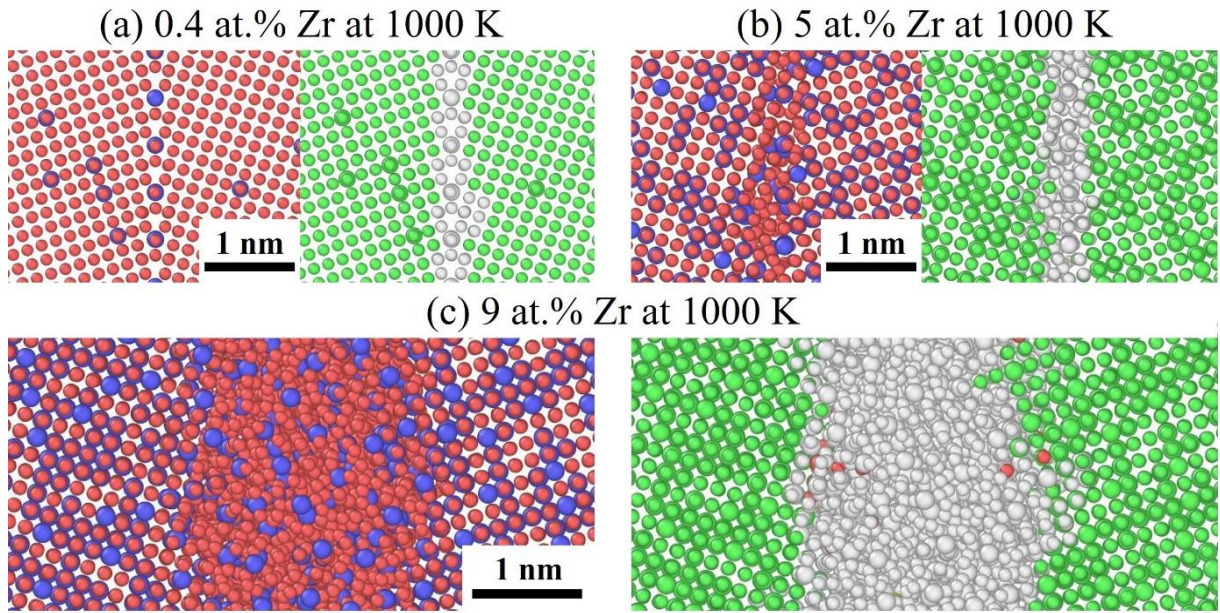

Fig. 12. The equilibrium chemical (left panel) and structural (right panel) information of the Σ 5(013) grain boundary in Ni doped with (a) 0.4 at.% Zr at 1000 K, (b) 5 at.% Zr at 1000 K, and (c) 9 at.% Zr at 1000 K.    In the panels on the left, Ni atoms are colored red and Zr atoms are colored blue. In the panels on the right, face centered cubic atoms are colored green, hexagonal close packed atoms red, body centered cubic atoms purple, icosahedral atoms yellow, and other atoms white.

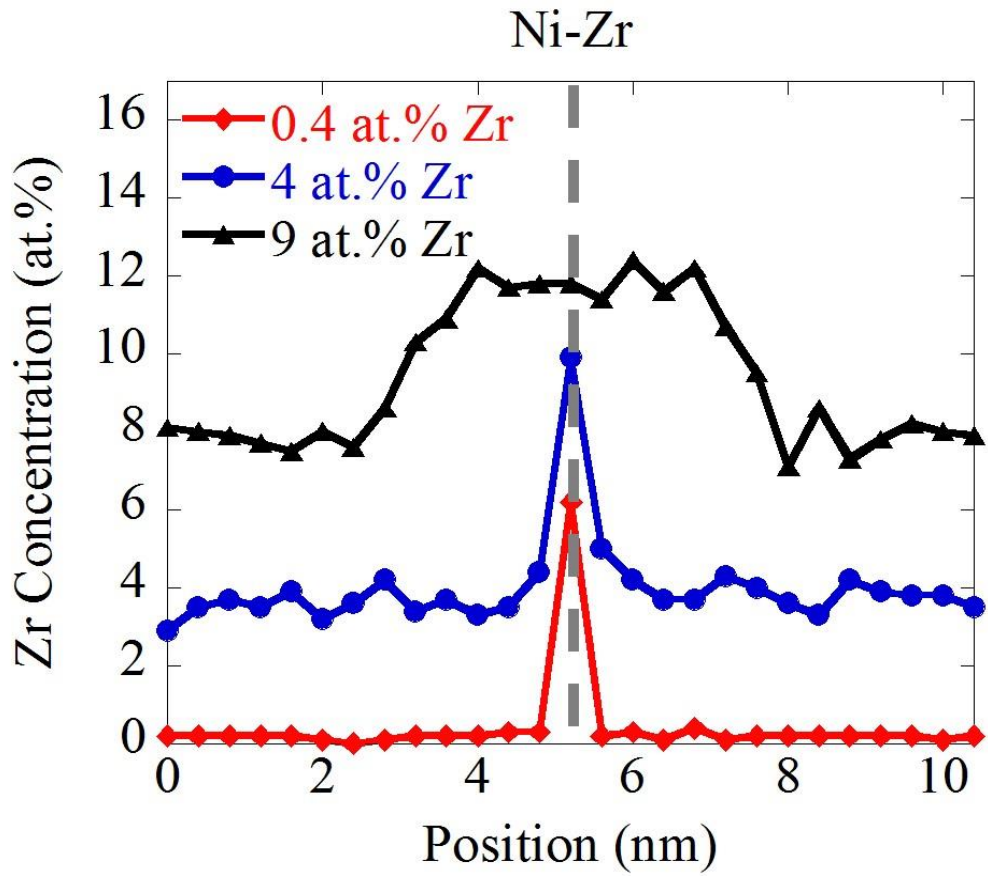

Fig. 13. Zr concentration profile across the grain boundary in Ni samples doped with 0.4 at.% Zr, 5 at.% Zr, and 9 at.% Zr at 1000 K.